# Minimum curvilinear automata with similarity attachment for network embedding and link prediction in the hyperbolic space

Alessandro Muscoloni[1] and Carlo Vittorio Cannistraci[1,2]

[1]Biomedical Cybernetics Group, Biotechnology Center (BIOTEC), Center for Molecular and Cellular Bioengineering (CMCB), Center for Systems Biology Dresden (CSBD), Department of Physics, Technische Universität Dresden, Tatzberg 47/49, 01307 Dresden, Germany; [2]Brain bio-inspired computing (BBC) lab, IRCCS Centro Neurolesi "Bonino Pulejo", Messina, Italy

## Abstract

The idea of minimum curvilinearity (MC) is that the hidden geometry of complex networks, in particular when they are sufficiently sparse, clustered, small-world and heterogeneous, can be efficiently navigated using the minimum spanning tree (MST), which is a greedy navigator. The local topological information drives the global geometrical navigation and the MST can be interpreted as a growing path that greedily maximizes local similarity between the nodes attached at each step by globally minimizing their overall distances in the network. This is also valid in absence of the network structure and in presence of only the nodes geometrically located over the network generative manifold in a high-dimensional space.

We know that random geometric graphs in the hyperbolic space are an adequate model for realistic complex networks: the explanation of this connection is that complex networks exhibit hierarchical, tree-like organization, and in turn the hyperbolic geometry is the geometry of trees. Here we show that, according to a mechanism that we define *similarity attachment*, the visited node sequence of a network automaton - which navigates the network with a growing MST in compliance with the minimum curvilinearity strategy - can efficiently approximate the nodes' angular coordinates in the hyperbolic disk, that actually represent an ordering of their similarities. This is a consequence of the fact that the MST, during its greedy growing process, at each step sequentially attaches the node most similar (less distant) to its own cohort.

Minimum curvilinear automata (MCA) displays embedding accuracy which seems superior to HyperMap-CN and inferior to coalescent embedding, however its link prediction performance on real networks is without precedent for methods based on the hyperbolic space. Finally, depending on the data structure used to build the MST, the MCA's time complexity can also approach a linear dependence from the number of edges.

# Introduction

The problem of embedding complex networks into the hyperbolic disk using only the unweighted topological information - especially when they are clustered, small-world and scale-free - has recently attracted attention in network science. Indeed, the computational solution of this inverse problem in physics of complex systems favors the application of network latent geometry techniques in disciplines dealing with big network data analysis including biology, medicine, and social science.

The first methods developed [1]–[3] are *generative-model-based* and were proposed exploiting the idea to infer the hyperbolic coordinates maximizing the likelihood that the network has been produced by a random geometrical generative model [1], that for instance can be the popularity-similarity-optimization (PSO) model [4]. More recently, different *model-free* methods have been presented [5]–[7] and even a hybrid one [8]. Among them, the one offering the best performance for correct angular coordinates inference seems a nonlinear-dimension-reduction-based method named coalescent embedding [7], which also has the virtue to reduce the computational time complexity of the approximated solution to $O(N^2)$.

Here, we present a *mechanistic-model* named minimum curvilinear automaton (MCA) which relies on a mechanism of node growth that we named *similarity attachment*, and that allows an efficient inference of the angular coordinates and embedding of networks in the hyperbolic disk. The revolutionary idea is that a local rule of node attachment implements a network automaton that grows according to a strategy of network navigation termed minimum curvilinearity. A spanning tree (not minimum) has been adopted also in a previous work by Kleinberg [9], in which a hyperbolic embedding is generated with the only purpose of maximizing the greedy routing. However, the way in which the spanning tree is exploited is completely different from the one we propose and the mechanism of similarity attachment was not present. The results discussed in the following section will provide evaluations of the performance of this new method in comparison to the previous ones both on synthetic and real networks.

# Results and Discussion

Figure 1 and Video 1 (link: https://youtu.be/xNzw8zCSnJI) show the core algorithmically idea of MCA. MCA are network automata growing according to a strategy named minimum curvilinearity. The idea of minimum curvilinearity (MC) is that the hidden geometry of complex networks that are in particular sufficiently sparse, clustered, small-world and heterogeneous can be efficiently navigated using the minimum spanning tree (MST), which is a greedy navigator. The local topological information drives the global geometrical navigation and the MST can be interpreted as a growing path that greedily maximizes local similarity between the nodes attached at each step by globally minimizing their overall distances in the network. This is also valid in absence of the network structure and in presence of only the nodes geometrically located over the network generative manifold in a high-dimensional space [10], [11]. We know that random geometric graphs in the hyperbolic space are an adequate model for realistic complex networks [4]: the explanation of this connection is that complex networks exhibit hierarchical, tree-like organization, and in turn the hyperbolic geometry is the geometry of trees [12]. Here we show that, according to a mechanism we define node similarity preferential attachment (in short: *similarity attachment*), the visited node sequence of a network automaton - which navigates the network with a growing MST in compliance with the minimum curvilinearity strategy - can efficiently approximate the nodes' angular coordinates in the hyperbolic disk, that actually represent an ordering of their similarities. This is a consequence of the fact that the MST, during its greedy growing process (for instance adopting the Prim's algorithm [13]), at each step sequentially attaches the node most similar (less distant) to its own cohort. This is indeed displayed in the Figure 1 where the MCA growth is represented at the initial, intermediated and final stage. The entire dynamic process of node angular ordering and network embedding is instead represented in the Video 1.

Figure 2A reports the detailed flow-chart of the algorithm for embedding in the hyperbolic disk that adopts MCA and that, for simplicity, we also call MCA. The network should be firstly pre-weighted in order to guide the MST growth according to a "good guess" of the edge weights that suggest the connectivity geometry. We propose two possible pre-weighting options based on a local rule named repulsion-attraction: RA1 and RA2 (the formulas are described in Figure 2). Then, the MCA is executed on the weighted network to provide a circular ordering of the nodes. We designed two variants: MCA1 (displayed in Figure 1 and Video 1), in which the new node attached at each step to the growing MST is located always on the same side of the growing circular ordering; MCA2 (not displayed), in which the new node is located on the side that is closer to the MST node to which it attaches (or in the same side as the previous step in case of tie). Given the circular ordering, the angular distances of nodes adjacent on the circumference of the disk can be either equidistantly adjusted (EA), or adjusted proportionally to the distances approximated by the respective RA-rule adopted to pre-weight the network (repulsion-attraction adjustment, RAA). Finally, the radial coordinates are assigned according to a widely used formula based on the PSO theory [2]. The detailed algorithmic steps are described in Suppl. Algorithm 1.

Figure 2B visually shows the result of the MCA algorithm when applied to a realistic (since similarly to many real complex networks displays clustering, small-worldness, scale-freeness and community structure) artificial network generated by the nPSO model [14].

Figure 3 and Suppl. Figures 1-4 report the results of the evaluations according to HD-correlation and GR-score (the metrics are described in the figure legends and Suppl. Methods, for further details refer to [7]). The best performing version of MCA (RA2-MCA1-RAA) is compared to HyperMap-CN and the two main types of coalescent embedding (one hierarchical-based ncMCE and one manifold-based ncISO). MCA displays an embedding accuracy (HD-correlation) that in general seems superior to HyperMap-CN and inferior to coalescent embedding. However, the network mappings of coalescent embedding are less navigable (lower GR-score) than the ones of MCA, which appears as the most robust over the different evaluations since it offers a good performance in nPSO networks with low temperature (which have high clustering and likely resemble real network topologies) both in embedding-accuracy and greedy-navigability. Remarkably, MCA is also the best method for link prediction on many diverse real networks of different size and topological characteristics (Table 1), on which its performance is without precedent for methods based on the hyperbolic space. A comparison between MCA1 and MCA2 is provided in Suppl. Figures 5-8 and Suppl. Table 1, highlighting no major differences.

Finally, depending on the type of minimum edge-weight data structure used to grow the MST, the MCA's time complexity can also approach a linear dependence from the number of edges. Efficient implementations of the Prim's algorithm use an adjacency list and a priority queue, and two different data structures have been mainly adopted as a priority queue: a binary heap, which leads to a time complexity of $O(E \cdot \log N)$ [15], and a Fibonacci heap, which brings the complexity down to $O(E + N \cdot \log N)$ [16]. The latter solution corresponds to $O(N \cdot \log N)$ for sufficiently sparse networks and to $O(E)$ for denser networks. An empirical analysis that compared also other data structures [17] (pairing heap, rank-relaxed heap, splay tree, d-ary heap) suggested that, although the worst-case time complexity is still $O(E + N \cdot \log N)$, a pairing heap might be a simpler and faster solution in practice, with the trade-off of using a little more storage than a binary heap.

To conclude, this article proposes the first evidence that simple mechanistic-models of network automata growing by similarity attachment (rather than the known popularity attachment [18]) can be very efficient, providing appreciable performance with relatively low time complexity. MCA could play an important role in solving computational problems of network geometry that lie at the interface between physics of complex systems and network science.

## Authors' contributions

The study was envisioned and the MCA algorithm was invented by C.V.C. The MCA algorithmic variant MCA2 that orders the nodes on both the sides of the circumference was invented by A.M. The code was written and the simulations were implemented by A.M. The display items were designed by C.V.C with the help of A.M., and they were realized by A.M. The main text of the article was written by C.V.C., the methods were written by A.M, and both the Authors revised all the article. The project was led by C.V.C.

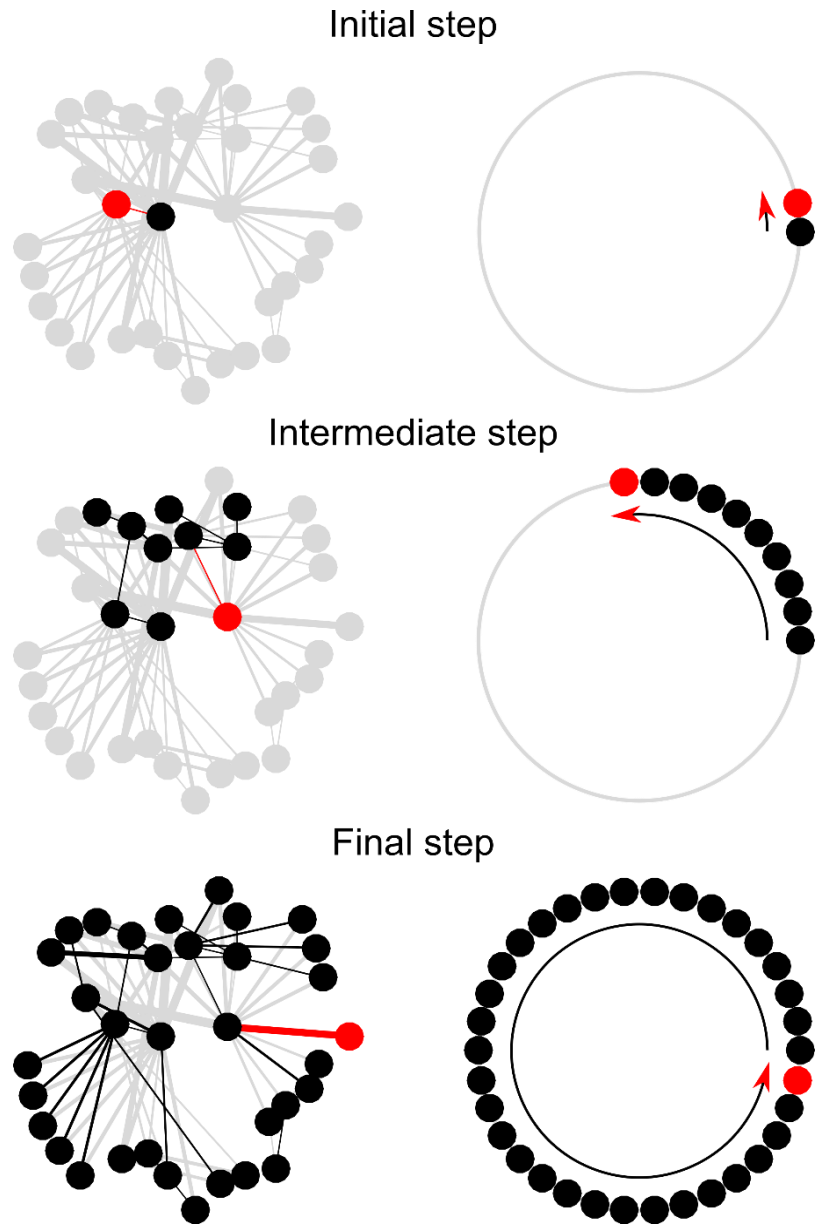

**Figure 1. The mechanism of** *similarity attachment*.
Visual representation of the mechanism that we define *similarity attachment*: the visited node sequence of the growing MST using the Prim's algorithm on a pre-weighted network represents an ordering of their similarities. The initial, intermediate and final steps are reported. The subplots on the left show the MST growing on the Karate network, pre-weighted using the RA1 rule: in black the nodes and links of the current tree, in red the node and link added at the next step, in grey the remaining nodes and links of the network. The line width is directly proportional to the link weight. The subplots on the right show the growing circular ordering of the nodes: in black the nodes in the current tree, in red the node added at the next step.

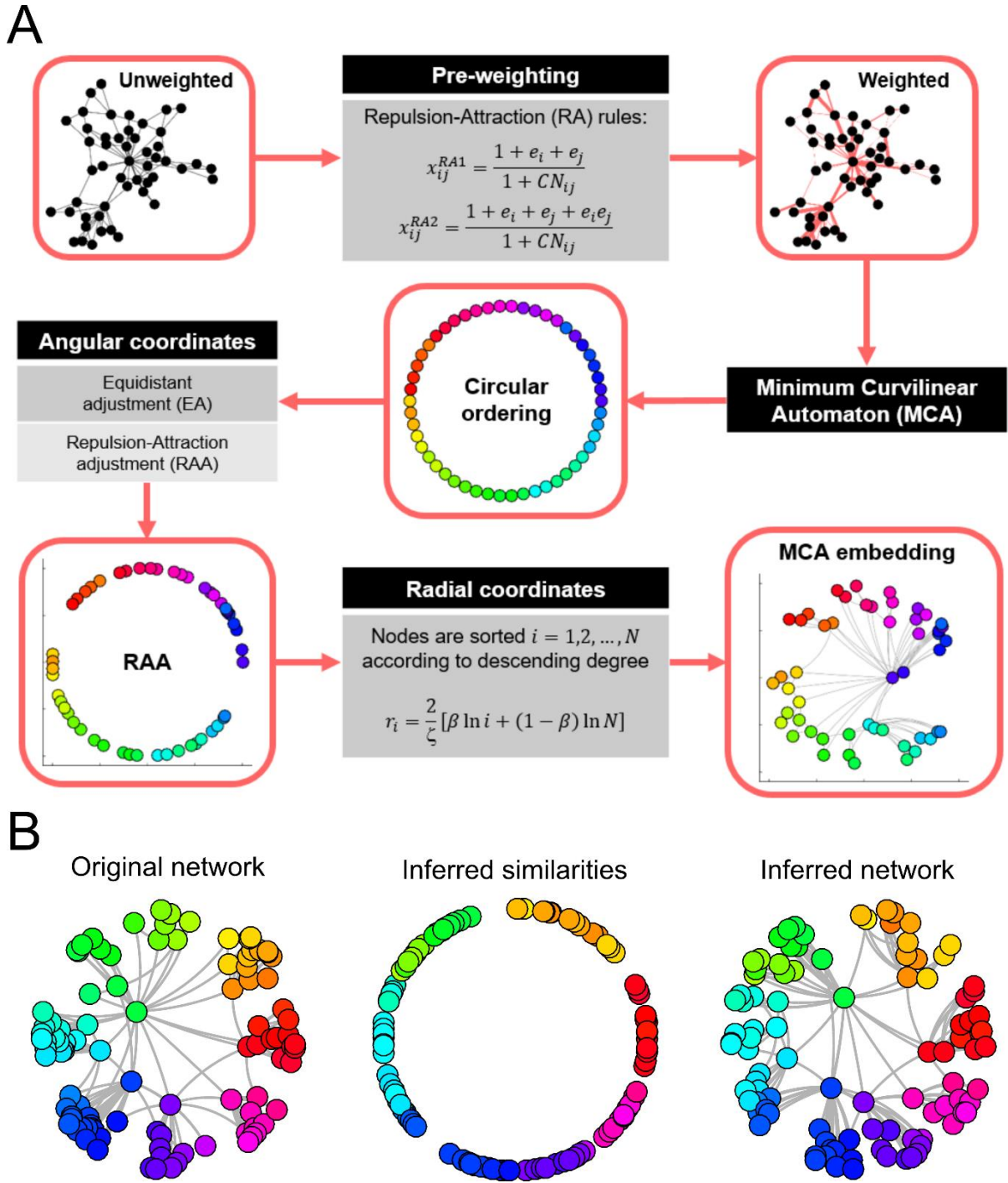

**Figure 2. Flow chart of the MCA embedding algorithm.**
**(A)** The algorithmic steps (greyscale squares) and the intermediate input/output (rounded red squares) of the MCA embedding algorithm are illustrated, reporting the possible variants. When the automaton is applied, it can be either MCA1 or MCA2. The example network has been generated by the PSO model ($N = 50$, $m = 2$, $T = 0$, $\gamma = 2.5$) and we applied RA1-MCA1-RAA. Description of the variables: $x_{ij}$ value of $(i,j)$ link in adjacency matrix $x$; $e_i$ external degree of node $i$ (links neither to $CN_{ij}$ nor to $j$); $CN_{ij}$ common neighbours of nodes $i$ and $j$; $N$ number of nodes; $\zeta = \sqrt{-K}$, we set $\zeta = 1$; $K$ curvature of the hyperbolic space; $\beta = \frac{1}{\gamma-1}$ popularity fading parameter; $\gamma$ exponent of power-law degree distribution. **(B)** The three subpanels show the original network generated with the nPSO model (N = 100, m = 2, T = 0, $\gamma$ = 2.5, C = 8), the similarities of the nodes inferred using the RA1-MCA1-RAA algorithm and the corresponding embedding of the network in the hyperbolic disk. In all the panels the nodes are colored according to their angular coordinates in the original network.

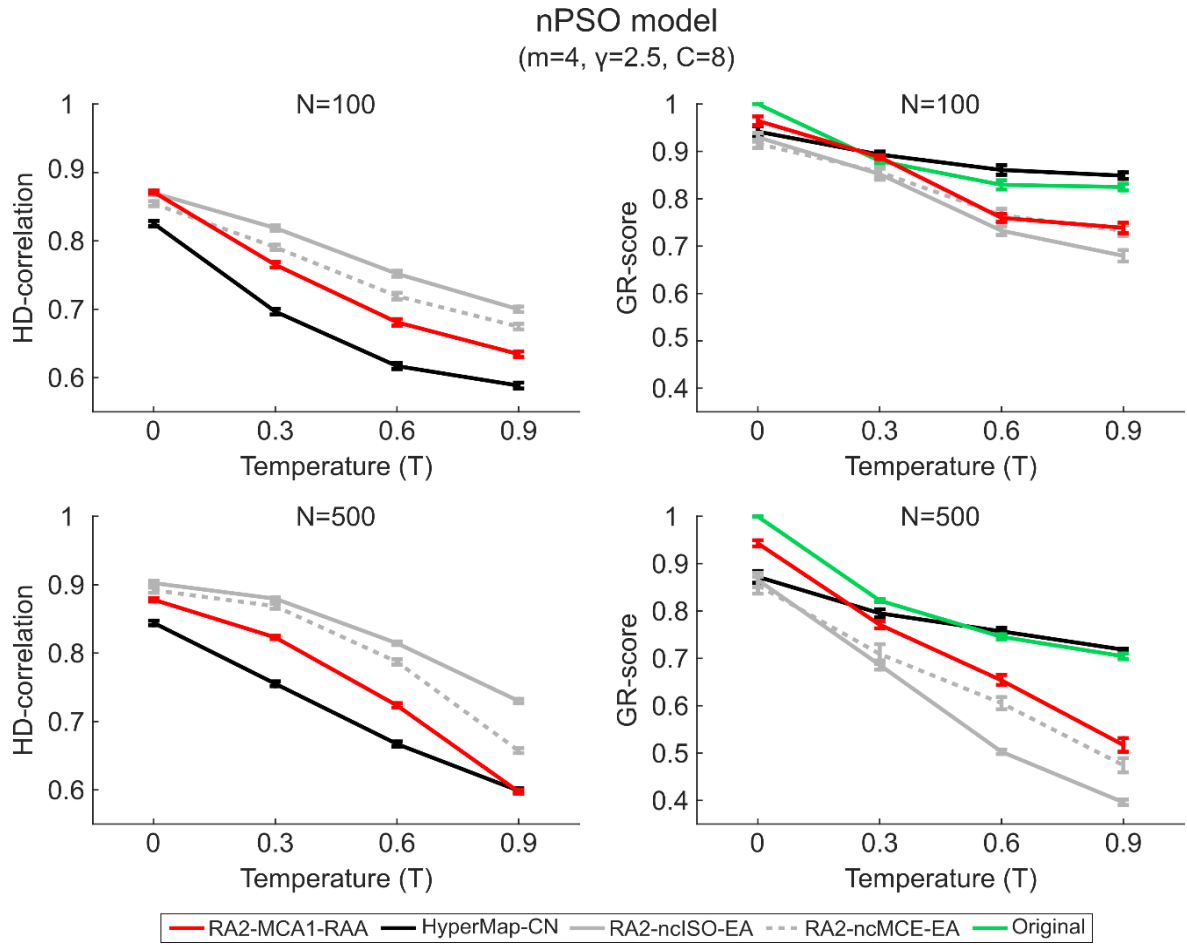

**Figure 3. MCA embedding and evaluation of nPSO networks.**
Synthetic networks have been generated using the nPSO model ($N = [100-500]$, $m = 4$, $T = [0, 0.3, 0.6, 0.9]$, $\gamma = 2.5$, $C = 8$, realizations = 100). For each network the hyperbolic embedding methods have been executed and the performance has been evaluated using the HD-correlation (Pearson correlation between all the pairwise hyperbolic distances of the nodes in the original and inferred network) and GR-score (greedy routing score), see Suppl. Methods for details [7]. The plots report for each parameter combination the mean and standard error over the 100 iterations. The methods compared are one MCA variant (RA2-MCA1-RAA), two main coalescent embedding techniques [7] (RA2-ncISO-EA and RA2-ncMCE-EA) and HyperMap-CN [3]. The GR-score evaluated on the original coordinates of the network is also reported.

|  | RA1 MCA1 RAA HSP | RA1 ncMCE RAA HSP | RA1 MCA1 RAA HD | RA1 ncMCE RAA HD | LPCS HSP | LPCS HD | LaBNE HM HSP | Hyper Map CN HSP | Hyper Map CN HD | LaBNE HM HD |
|---|---|---|---|---|---|---|---|---|---|---|
| Mouse neural | **0.12** | 0.11 | 0.04 | 0.03 | 0.03 | 0.05 | 0.06 | 0.02 | 0.03 | 0.05 |
| Karate | **0.15** | 0.11 | 0.07 | 0.03 | 0.09 | 0.08 | 0.05 | 0.11 | 0.14 | 0.03 |
| Dolphins | **0.15** | 0.09 | 0.10 | 0.05 | 0.06 | 0.05 | 0.05 | 0.04 | 0.01 | 0.01 |
| Macaque neural | **0.60** | 0.46 | 0.35 | 0.25 | 0.22 | 0.19 | 0.16 | 0.28 | 0.13 | 0.09 |
| Polbooks | **0.12** | 0.10 | 0.11 | 0.09 | 0.09 | 0.07 | 0.07 | 0.06 | 0.06 | 0.04 |
| ACM2009 contacts | **0.25** | 0.22 | 0.15 | 0.14 | 0.09 | 0.09 | 0.11 | 0.08 | 0.09 | 0.09 |
| Football | 0.30 | 0.27 | **0.31** | 0.26 | 0.22 | 0.21 | 0.08 | 0.15 | 0.18 | 0.01 |
| Physicians innovation | **0.04** | **0.04** | **0.04** | **0.04** | 0.03 | 0.03 | 0.03 | 0.03 | 0.03 | 0.01 |
| Manufacturing email | **0.41** | 0.29 | 0.16 | 0.15 | 0.12 | 0.09 | 0.13 | 0.34 | 0.08 | 0.09 |
| Littlerock foodweb | **0.26** | 0.18 | 0.08 | 0.03 | 0.10 | 0.10 | 0.07 | 0.06 | 0.03 | 0.05 |
| Jazz | **0.35** | 0.29 | 0.30 | 0.18 | 0.15 | 0.08 | 0.09 | 0.15 | 0.06 | 0.02 |
| Residence hall friends | **0.15** | **0.15** | 0.13 | 0.12 | 0.08 | 0.05 | 0.05 | 0.07 | 0.07 | 0.01 |
| Haggle contacts | **0.56** | 0.55 | 0.44 | 0.47 | 0.52 | 0.42 | 0.26 | 0.26 | 0.00 | 0.09 |
| Worm nervoussys | **0.11** | 0.08 | 0.05 | 0.04 | 0.04 | 0.03 | 0.03 | 0.02 | 0.03 | 0.01 |
| Netsci | 0.21 | 0.18 | **0.31** | 0.23 | 0.09 | 0.11 | 0.08 | 0.10 | 0.13 | 0.01 |
| Infectious contacts | **0.25** | 0.24 | 0.24 | 0.19 | 0.13 | 0.05 | 0.10 | 0.11 | 0.04 | 0.02 |
| Flightmap | **0.54** | 0.52 | 0.46 | 0.44 | 0.48 | 0.44 | 0.28 | 0.11 | 0.02 | 0.23 |
| Email | **0.10** | 0.07 | 0.06 | 0.03 | 0.04 | 0.02 | 0.02 | 0.02 | 0.01 | 0.00 |
| Polblog | **0.15** | 0.12 | 0.12 | 0.09 | 0.07 | 0.05 | 0.05 | 0.04 | 0.03 | 0.01 |
| **mean precision** | **0.25** | 0.21 | 0.19 | 0.15 | 0.14 | 0.12 | 0.09 | 0.11 | 0.06 | 0.05 |
| **mean ranking** | **1.26** | 2.50 | 3.24 | 5.00 | 5.29 | 6.58 | 6.89 | 7.03 | 8.03 | 9.18 |

**Table 1. Precision evaluation of link-prediction on small-size real networks.**
For each network 10% of links have been randomly removed (100 realizations), the hyperbolic embedding methods have been executed and the non-observed links in these reduced networks have been ranked either by increasing hyperbolic distance (HD) between the nodes, or by increasing hyperbolic shortest path (HSP), computed as sum of the HD over the shortest path between the nodes [11]. In order to evaluate the performance, the precision is computed as the percentage of removed links among the top-$r$ in the ranking, where $r$ is the total number of links removed. The table reports for each network the mean precision over the 100 iterations, the mean precision over the entire dataset and the mean precision-ranking [19]. For each network the best performance is highlighted in bold. A description of the networks is provided as Supplementary Information. The methods compared are the best MCA variant (RA1-MCA1-RAA), the best coalescent embedding technique [7] (RA1-ncMCE-RAA), HyperMap-CN [3], LPCS [5] and LaBNE-HM [8].

# Supplementary Information

## Methods

**Minimum curvilinear automata (MCA)**

MCA are network automata growing according to a strategy named minimum curvilinearity. The idea of minimum curvilinearity (MC) is that the hidden geometry of complex networks that are in particular sufficiently sparse, clustered, small-world and heterogeneous can be efficiently navigated using the minimum spanning tree (MST), which is a greedy navigator. During each step of its greedy growing process (adopting the Prim's algorithm [1]), the MST sequentially attaches the node most similar (less distant) to its own cohort. Since the nodes angular coordinates in the hyperbolic disk actually represent an ordering of their similarities, they can be efficiently approximated by the visited node sequence of the MST.

The algorithm proceeds according to the following steps:

(1) Since in unweighted networks all the spanning trees are minimum, the inference of the similarities can be improved by pre-weighting the network with a convenient strategy that suggests the geometrical distances of the connections between the nodes [2]. We propose two possible pre-weighting options based on a local rule named repulsion-attraction (RA). The rule exploits only local information and is based on the idea that adjacent nodes with a high external degree (where the external degree is computed considering the number of neighbours not in common) should be geometrically far because they represent hubs without neighbours in common, which - according to the theory of navigability of complex networks presented by Boguñá et al. [3] - tend to dominate geometrically distant regions: this is the repulsive part of the rule. On the contrary, adjacent nodes that share a high number of common neighbours should be geometrically close because most likely they share many similarities: this is the attractive part of the rule.

The two mathematical formulations are the following:

$$x_{ij}^{RA1} = \frac{1 + e_i + e_j}{1 + CN_{ij}}$$

$$x_{ij}^{RA2} = \frac{1 + e_i + e_j + e_i e_j}{1 + CN_{ij}}$$

where $x_{ij}$ is the value of $(i,j)$ link in adjacency matrix $x$; $e_i$ is the external degree of node $i$ (links neither to $CN_{ij}$ nor to $j$); $CN_{ij}$ are the common neighbours of nodes $i$ and $j$.

(2) The Prim's algorithm [1] is executed over the pre-weighted network. At step $t = 1$, it initializes the MST with one node. We tested both using the highest degree node or a random node and no significant differences have been detected, therefore we show only the results for the deterministic variant, with the advantage of reproducibility. At each next step $t = 2 \dots N$, it finds the edge of minimum weight between nodes already in the tree and nodes not yet in the tree. Such edge, and the related node not yet in the tree, are attached to the tree. The sequence in which the nodes are introduced in the growing MST represents a circular ordering of their similarities.

In our tests we designed two variants. In the first variant (MCA1) the sequence grows only in one direction and therefore the new node attached at each step to the growing MST is located always on the same side of the growing circular ordering. In the second variant (MCA2) the sequence can grow in both the directions and the new node is located on the side that is closer to the MST node to which it attaches (or in the same side as the previous step in case of tie). MCA2 offers a slightly higher performance on synthetic networks (Suppl. Figures 5-8), but no improvements on real networks (Suppl. Table 1). Therefore, in the main article we decided to present only the algorithmically simplest variant MCA1.

(3) The nodes are arranged over the circumference of the disk according to the circular ordering obtained at the previous step. The angular distances between adjacent nodes can be either equidistantly adjusted (EA), or adjusted proportionally to the distances approximated by the respective RA-rule adopted to pre-weight the network (repulsion-attraction adjustment, RAA).

In case of equidistant adjustment, the angular coordinates are assigned as follows:

$$\theta_i = \frac{2\pi}{N}(t_i - 1)$$

where $N$ is the number of nodes in the network, $\theta_i$ is the angular coordinate of the node $i = 1 \ldots N$ and $t_i \in [1, N]$ is the step in which the node $i$ has been added to the MST (its position in the ordered sequence).

In case of repulsion-attraction adjustment, the angular coordinates are assigned as follows. Given the (circular) sequence of nodes $s_1, s_2, \ldots s_N$, the repulsion-attraction is computed between adjacent nodes, creating a vector of $N$ weights.

$$w_1 = RA(s_1, s_2); \ w_2 = RA(s_2, s_3); \ \ldots ; \ w_{N-1} = RA(s_{N-1}, s_N); \ w_N = RA(s_N, s_1)$$

Then, the weights are normalized so that their sum is $2\pi$.

$$w_j^* = \frac{w_j}{\sum_{k=1}^{N} w_k} * 2\pi$$

Finally, the angular coordinates are assigned.

$$\theta_i = \sum_{j=1}^{t_i - 1} w_j^*$$

(4) A maximum likelihood estimation of the sequence according to which nodes appeared in the network indicates that the higher the degree of the node, the earlier it appeared [4]. Therefore, nodes are sorted according to descending degree and the radial coordinate of the $i$-th node in the set is computed according to:

$$r_i = \frac{2}{\zeta}[\beta \ln i + (1 - \beta) \ln N] \quad i = 1, 2, \ldots, N$$

$N$ number of nodes; $\zeta = \sqrt{-K}$, we set $\zeta = 1$; $K$ curvature of the hyperbolic space; $\beta = \frac{1}{\gamma - 1}$ popularity fading parameter; $\gamma$ exponent of power-law degree distribution.

The exponent $\gamma$ of the power-law degree distribution has been fitted using the MATLAB script '*plfit.m*', according to a procedure described by Clauset et al. [5] and released at http://www.santafe.edu/~aaronc/powerlaws/. If a network has $\gamma < 2 \rightarrow \beta > 1$, which is out of the domain $\beta \in (0, 1]$ imposed by the PSO model, a few of the highest degree nodes should obtain a radial coordinate that indicates a popularity higher than the maximum popularity allowed ($r = 0$), but since it is not possible due to the previous equation, the radial coordinate degenerates and becomes negative. Hence, for these nodes we set $r = 0$.

**Coalescent embedding**

The expression coalescent embedding refers to a topological-based machine learning class of algorithms that exploits nonlinear unsupervised dimensionality reduction to infer the nodes angular coordinates in the hyperbolic space [2]. The techniques are able to perform a fast and accurate mapping of a network in the 2D hyperbolic disk and in the 3D hyperbolic sphere.

The first step of the algorithm for a 2D embedding consists in pre-weighting the network in order to suggest geometrical distances between connected nodes, since it has been shown that improves the mapping accuracy [2]. Two topological-based pre-weighting rules have been proposed, repulsion-attraction (RA) and edge-betweenness-centrality (EBC), respectively using local (RA) and global (EBC) topological information [2].

Given the weighted network, the second step consists in performing the nonlinear dimensionality reduction. Two different kinds of machine learning approaches can be used, manifold-based (LE, ISO, ncISO) or Minimum-Curvilinearity-based (MCE, ncMCE). The details about which dimensions of the embedding should be considered are provided in the original publication [2].

In order to assign the angular coordinates in the 2D embedding space, either a circular adjustment or an equidistant angular adjustment (EA) can be considered. The circular adjustment for the manifold-based approaches consists in exploiting directly the polar coordinates of the 2D reduced space, whereas for the Minimum-Curvilinearity-based it consists in rearranging the node points on the circumference following the same ordering of the 1D reduced space and proportionally preserving the distances. Using the equidistant angular adjustment, instead, the nodes are equidistantly arranged on the circumference, which might help to correct for short-range angular noise present in the embedding [2]. In this work, the same repulsion-attraction adjustment (RAA) introduced for the MCA algorithm has been adopted also for the coalescent embedding techniques.

The radial coordinates are assigned as described for the MCA method.

The code is available at https://github.com/biomedical-cybernetics/coalescent_embedding.

**HyperMap-CN**

HyperMap [4] is a method to map a network into the hyperbolic space based on Maximum Likelihood Estimation (MLE). For sake of clarity, the first algorithm for MLE-based network embedding in the hyperbolic space is not HyperMap, but to the best of our knowledge is the algorithm proposed by Boguñá et al. in [6]. HyperMap is basically an extension of that method applied to the PSO model [7]. It replays the hyperbolic growth of the network and at each time step $i$ it finds the coordinates of the added node $i$ by maximizing the likelihood that the network was produced by the E-PSO model [4]. According to the MLE procedure, the nodes are added in decreasing order of degree. The radial coordinates depend on the time step $i$ and on the exponent $\gamma$ of the power-law degree distribution. The angular coordinates, instead, are assigned by maximizing a likelihood function $L_{i,L}$, with the aim of mapping connected nodes at a low hyperbolic distance and disconnected nodes at a high hyperbolic distance. The maximization is done by numerically trying different angular coordinates in steps of $2\pi/N$ and choosing the one that leads to the biggest $L_{i,L}$.

HyperMap-CN [8] is a further development of HyperMap, where the inference of the angular coordinates is not performed anymore maximizing the likelihood $L_{i,L}$, based on the connections and disconnections of the nodes, but using another local likelihood $L_{i,CN}$, based on the number of common neighbours between each node $i$ and the previous nodes $j < i$ at final time. Here the hybrid model has been used, a variant of the method in which the likelihood $L_{i,CN}$ is only adopted for the high degree nodes and $L_{i,L}$ for the others, yielding a shorter running time. Furthermore, a speed-up heuristic and corrections steps can be applied. The speed-up can be achieved by getting an initial estimate of the angular coordinate of a node $i$ only considering the previous nodes $j < i$ that are $i$'s neighbours, the maximum likelihood estimation is then performed only looking at an interval around this initial estimate. Correction steps can be used at predefined times $i$: each existing node $j < i$ is visited and with the knowledge of the rest of the coordinates the angle of $j$ is updated to the value that maximizes the likelihood $L_{j,L}$. The C++ implementation of the method has been released by the authors at the website https://bitbucket.org/dk-lab/2015_code_hypermap. In our simulations, neither correction steps nor speed-up heuristic have been used. The input parameter $\gamma$ has been fitted as described for the MCA method. Based on the assumption that the clustering coefficient decreases almost linearly with the network temperature, until it is 0 for $T = 1$ [7], the following procedure has been proposed [9] in order to choose the input parameter $T$ (temperature): (1) ten PSO synthetic networks are generated with $T = 0$ and the same parameters $N$, $m$ and $\gamma$ as the given network; (2) the clustering coefficient of the ten networks is averaged and used as y-intercept, while the point ($T = 1$, *clustering* = 0) is used as x-intercept; (3) from the equation of this line and the clustering coefficient of the given network, we can estimate its temperature $T$ [9]. Although this procedure is possible when few networks have to be embedded, it becomes too time consuming in wide simulations. Therefore, in all our synthetic and real mappings we used a default value $T = 0.1$.

**LaBNE+HM**

Laplacian-based Network Embedding (LaBNE) [10] is a hyperbolic embedding technique that exploits nonlinear dimensionality reduction to infer the nodes angular coordinates in the hyperbolic space. Indeed, it applies Laplacian Eigenmaps (LE) to the unweighted adjacency matrix in order to embed the network in a two-dimensional space, and exploits the angular coordinates of the nodes in this reduced space as angular coordinates for the hyperbolic disk. The radial coordinates are assigned as described for the MCA method. We let notice that this method is a particular case of coalescent embedding [2], in which no pre-weighting rules are adopted, the LE nonlinear dimensionality reduction technique is performed, and the circular adjustment is used for assigning the angular coordinates.

LaBNE+HM [9] is a hybrid approach that combines the higher computational speed of LaBNE with the higher mapping accuracy of HyperMap. At first, LaBNE is executed in order to obtain a preliminary embedding of the network in the hyperbolic space. As second, HyperMap refines the embedding trying to find the best angular coordinate of each node, maximizing the likelihood $L_{i,L}$ within a window around the initial LaBNE estimate. The input parameters $\gamma$ and $T$ has been set as for HyperMap-CN, whereas the search window has been fixed to $\pi/4$.
The R implementation is available at https://github.com/galanisl/NetHypGeom.

**LPCS**

Link Prediction with Community Structure (LPCS) [11] is a hyperbolic embedding technique that consists of the following steps: (1) Detect the hierarchical organization of communities. (2) Order the top-level communities starting from the one that has the largest number of nodes and using the Community Intimacy index, which takes into account the proportion of edges within and between communities. (3) Recursively order the lower level communities based on the order of the higher-level communities, until reaching the bottom level in the hierarchy. (4) Assign to every bottom-level community an angular range of size proportional to the nodes in the community, in order to cover the complete circle with non-overlapping angular ranges. Sample the angular coordinates of the nodes uniformly at random within the angular range of the related bottom-level community. (5) Assign the radial coordinates as described for the MCA method.

The LPCS code firstly takes advantage of the R function *multilevel.community* for detecting the hierarchy of communities, an implementation of the Louvain method [12] available in the *igraph* package [13], while the following embedding steps have been implemented in MATLAB.

**Greedy routing**

An important characteristic that can be studied in a network embedded in a geometrical space is its navigability. The network is considered navigable if the greedy routing (GR) performed using the node coordinates in the geometrical space is efficient [3]. In the GR, for each pair of nodes $i$ and $j$, a packet is sent from $i$ with destination $j$. Each node knows only the address of its neighbours and the address of the destination $j$, which is written in the packet. The address of a node is represented by its coordinates in the geometrical space. In the GR procedure adopted [4], the nodes are located in the hyperbolic disk and at each hop the packet is forwarded from the current node to its neighbour closest to the destination, meaning at the lowest hyperbolic distance. The packet is dropped when this neighbour is the same from which the packet has been received at the previous hop, since a loop has been generated. In order to evaluate the efficiency of the GR, two metrics are usually taken into account: the percentage of successful paths and the average hop-length of the successful paths [3]. The first one indicates the proportion of packets that are able to reach their destinations - the higher the better - whereas the second one indicates if the successful packets require on average a short path to reach the destination - the lower the better. In order to compare the performance of methods in a unique way while taking both of the metrics into account, a GR-score has been introduced [2] and it is computed as follows:

$$GR_{score} = \frac{\sum_{i=1}^{N}\sum_{j=1, j \neq i}^{N} \frac{sp_{ij}}{p_{ij}}}{N*(N-1)}$$

Where $i$ and $j$ are two within the set of $N$ nodes, $sp_{ij}$ is the shortest path length from $i$ to $j$ and $p_{ij}$ is the GR path length from $i$ to $j$. The ratio $\frac{sp_{ij}}{p_{ij}}$ assumes values in the interval [0, 1]. When the greedy routing is unsuccessful the path length is infinite and therefore the ratio is 0, which represents the worst case. When the greedy routing is successful the path length is greater than 0 and tends to 1 as the path length tends to the shortest path length, becoming 1 in the best case. The GR-score is the average of this ratio over all the node pairs.

The code is available at https://github.com/biomedical-cybernetics/coalescent_embedding.

**Hyperbolic shortest paths for link prediction**

A practical application of the network embedding in the hyperbolic space is the prediction of missing or future links. Since according to the original PSO model the connection probability is inversely related to the hyperbolic distance (HD) [7], the methods based on hyperbolic embedding that have been developed so far used the HD in order to assign likelihood scores to the non-observed links [4], [8], [11]. In particular, the lower the HD between two disconnected nodes, the higher the likelihood that the link is missing or that it will appear in the future. However, it has been shown for the Euclidean space that computing the distances over the topology (as shortest paths in the network weighted by the Euclidean distances), rather than as direct distances in the geometrical space, can improve the discrimination between good and bad candidate links and therefore the link prediction performance [14]. The same theory has been adopted for the link prediction in the hyperbolic space: the network is weighted using the HD and the hyperbolic shortest paths (HSP) are computed as sum of the HD over the shortest path between each pair of nodes. The lower the HSP between two disconnected nodes, the higher the likelihood that the link is missing or that it will appear in the future.

**Datasets**

**Generation of synthetic networks using the PSO model**

The Popularity-Similarity-Optimization (PSO) model [7] is a generative network model recently introduced in order to describe how random geometric graphs grow in the hyperbolic space. In this model the networks evolve optimizing a trade-off between node popularity, abstracted by the radial coordinate, and similarity, represented by the angular coordinate distance, and they exhibit many common structural and dynamical characteristics of real networks.

The model has four input parameters:
- $m > 0$, which is equal to half of the average node degree;
- $\beta \in (0, 1]$, defining the exponent $\gamma = 1 + 1/\beta$ of the power-law degree distribution;
- $T \geq 0$, which controls the network clustering; the network clustering is maximized at $T = 0$, it decreases almost linearly for $T = [0,1)$ and it becomes asymptotically zero if $T > 1$;
- $\zeta = \sqrt{-K} > 0$, where $K$ is the curvature of the hyperbolic plane. Since changing $\zeta$ rescales the node radial coordinates and this does not affect the topological properties of networks [7], we considered $K = -1$.

Building a network of $N$ nodes on the hyperbolic disk requires the following steps:

(1) Initially the network is empty;
(2) At time $i = 1, 2, ..., N$ a new node $i$ appears with radial coordinate $r_i = 2ln(i)$ and angular coordinate $\theta_i$ uniformly sampled in $[0, 2\pi]$; all the existing nodes $j < i$ increase their radial coordinates according to $r_j(i) = \beta r_j + (1 - \beta) r_i$ in order to simulate popularity fading;
(3) If $T = 0$, the new node connects to the $m$ hyperbolically closest nodes; if $T > 0$, the new node picks a randomly chosen existing node $j < i$ and, given that it is not already connected to it, it connects to it with probability

$$p(i,j) = \frac{1}{1 + \exp\left(\frac{h_{ij} - R_i}{2T}\right)}$$

repeating the procedure until it becomes connected to $m$ nodes.
Note that

$$R_i = r_i - 2\ln\left[\frac{2T(1 - e^{-(1-\beta)\ln(i)})}{\sin(T\pi) m (1 - \beta)}\right]$$

is the current radius of the hyperbolic disk, and
$$h_{ij} = arccosh(\cosh r_i \cosh r_j - \sinh r_i \sinh r_j \cos \theta_{ij})$$
is the hyperbolic distance between node $i$ and node $j$, where
$$\theta_{ij} = \pi - \left|\pi - |\theta_i - \theta_j|\right|$$
is the angle between these nodes.
(4) The growing process stops when $N$ nodes have been introduced.

**Generation of synthetic networks using the nonuniform PSO (nPSO) model**

The nonuniform PSO (nPSO) model [15] is a variation of the PSO model introduced in order to confer to the generated networks an adequate community structure, which is lacking in the original model. Since the connection probabilities are inversely proportional to the hyperbolic distances, a uniform distribution of the nodes over the hyperbolic disk does not create agglomerates of nodes that are concentrated on angular sectors and that are more densely connected between each other than with the rest of the network. A nonuniform distribution, instead, allows to do it by generating heterogeneity in angular node arrangement. In particular, without loss of generality, we will concentrate on the Gaussian mixture distribution, which we consider suitable for describing how to build a nonuniform distributed sample of nodes along the angular coordinates of a hyperbolic disk, with communities that emerge in correspondence of the different Gaussians.
A Gaussian mixture distribution is characterized by the following parameters [16]:
- $C > 0$, which is the number of components, each one representative of a community;
- $\mu_{1...C} \in [0, 2\pi]$, which are the means of every component, representing the central locations of the communities in the angular space;
- $\sigma_{1...C} > 0$, which are the standard deviations of every component, determining how much the communities are spread in the angular space; a low value leads to isolated communities, a high value makes the adjacent communities to overlap;
- $\rho_{1...C}$ ($\sum_i \rho_i = 1$), which are the mixing proportions of every component, determining the relative sizes of the communities.

Given the parameters of the PSO model $(m, \beta, T)$ and the parameters of the Gaussian mixture distribution $(C, \mu_{1...C}, \sigma_{1...C}, \rho_{1...C})$, the procedure to generate a network of $N$ nodes is the same described in the section for the uniform case, with the only difference that the angular coordinates of the nodes are not sampled uniformly but according to the Gaussian mixture distribution. Note that, although the means of the components are located in $[0, 2\pi]$, the sampling of the angular coordinate $\theta$ can fall out of this range. In this case, it has to be shifted within the original range, as follows:

- If $\theta < 0 \rightarrow \theta = 2\pi - modulo(|\theta|, 2\pi)$
- If $\theta > 2\pi \rightarrow \theta = modulo(\theta, 2\pi)$

Although the parameters of the Gaussian mixture distribution allow for the investigation of disparate scenarios, as a first case of study we focused on the most straightforward setting. For a given number of components $C$, we considered their means equidistantly arranged over the angular space, the same standard deviation and equal mixing proportions:

- $\mu_i = \frac{2\pi}{C} * (i - 1) \quad i = 1 \ldots C$
- $\sigma_1 = \sigma_2 = \ldots = \sigma_C = \sigma$
- $\rho_1 = \rho_2 = \ldots = \rho_C = \frac{1}{C}$

In particular, in our simulations we fixed the standard deviation to 1/6 of the distance between two adjacent means $\left(\sigma = \frac{1}{6} * \frac{2\pi}{C}\right)$, which allowed for a reasonable isolation of the communities. The community memberships are assigned considering for each node the component whose mean is the closest in the angular space.

The code is available at https://github.com/biomedical-cybernetics/nPSO_model.

**Real networks**

The real networks have been transformed into undirected and unweighted, self-loops have been removed and the largest connected component has been considered.

*Mouse neural* (N=18, E=37): in-vivo single neuron connectome that reports mouse primary visual cortex (layers 1, 2/3 and upper 4) synaptic connections between neurons [17].

*Karate* (N=34, E=78): social network of a university karate club collected by Wayne Zachary in 1977. Each node represents a member of the club and each edge represents a tie between two members of the club [18].

*Dolphins* (N=62, E=159): a social network of bottlenose dolphins. The nodes are the bottlenose dolphins (genus Tursiops) of a bottlenose dolphin community living off Doubtful Sound, a fjord in New Zealand. An edge indicates a frequent association. The dolphins were observed between 1994 and 2001 [19].

*Macaque neural* (N=94, E=1515): a macaque cortical connectome, assembled in previous studies in order to merge partial information obtained from disparate literature and database sources [20].

*Polbooks* (N=105, E=441): nodes represent books about US politics sold by the online bookseller Amazon.com. Edges represent frequent co-purchasing of books by the same buyers, as indicated by the "customers who bought this book also bought these other books" feature on

Amazon. The network was compiled by V. Krebs and is unpublished, but can found at http://www-personal.umich.edu/~mejn/netdata/.

*ACM2009_contacts* (N=113, E=2196): network of face-to-face contacts (active for at least 20 seconds) of the attendees of the ACM Conference on Hypertext and Hypermedia 2009 [21].

*Football* (N=115, E=613): network of American football games between Division IA colleges during regular season Fall 2000 [22].

*Physicians innovation* (N=117, E=465): the network captures innovation spread among physicians in the towns in Illinois, Peoria, Bloomington, Quincy and Galesburg. The data was collected in 1966. A node represents a physician and an edge between two physicians shows that the left physician told that the right physician is his friend or that he turns to the right physician if he needs advice or is interested in a discussion [23].

*Manufacturing email* (N=167, E=3250): email communication network between employees of a mid-sized manufacturing company [24].

*Littlerock foodweb* (N=183, E=2434): food web of Little Rock Lake, Wisconsin in the United States of America. Nodes are autotrophs, herbivores, carnivores and decomposers; links represent food sources [25].

*Jazz* (N=198, E=2742): collaboration network between Jazz musicians. Each node is a Jazz musician and an edge denotes that two musicians have played together in a band. The data was collected in 2003 [26].

*Residence hall friends* (N=217, E=1839): friendship network between residents living at a residence hall located on the Australian National University campus [27].

*Haggle contacts* (N=274, E=2124): contacts between people measured by carried wireless devices. A node represents a person and an edge between two persons shows that there was a contact between them [28].

*Worm nervous* (N=297, E=2148): a C. *Elegans* connectome representing synaptic interactions between neurons [29].

*Netsci* (N=379, E=914): a co-authorship network of scientists working on networks science [30].

*Infectious contacts* (N=410, E=2765): network of face-to-face contacts (active for at least 20 seconds) of people during the exhibition INFECTIOUS: STAY AWAY in 2009 at the Science Gallery in Dublin [21].

*Flightmap* (N=456, E=37947): a network of flights between American and Canadian cities [31].

*Email* (N=1133, E=5451): email communication network at the University Rovira i Virgili in Tarragona in the south of Catalonia in Spain. Nodes are users and each edge represents that at least one email was sent [32].

*Polblog* (N=1222, E=16714): a network of front-page hyperlinks between blogs in the context of the 2004 US election. A node represents a blog and an edge represents a hyperlink between two blogs [33].

Most of the networks in the dataset can be downloaded from the Koblenz Network Collection at http://konect.uni-koblenz.de.

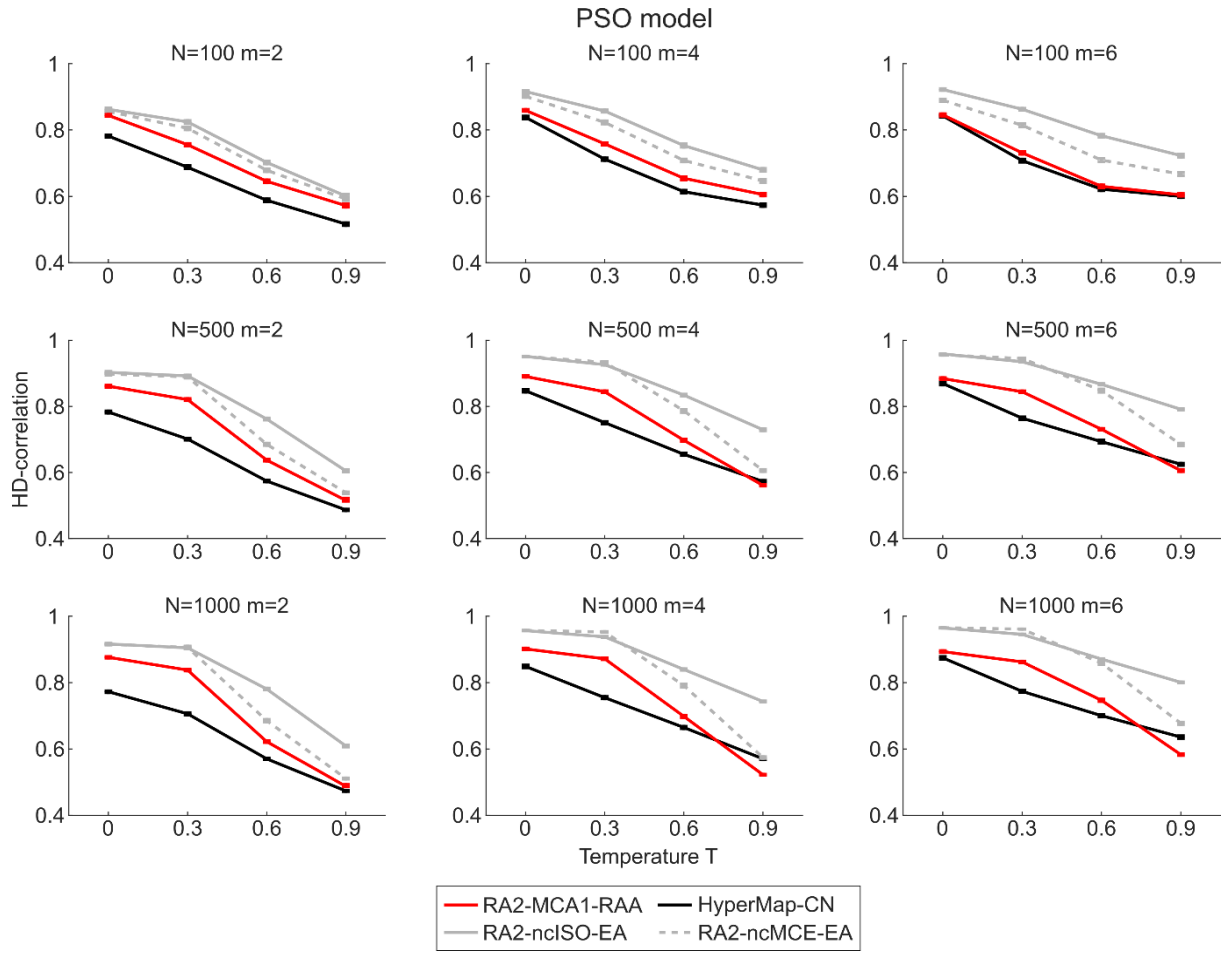

**Suppl. Figure 1. HD-correlation on PSO synthetic networks.**
Synthetic networks have been generated using the PSO model with parameters $\gamma = 2.5$ (power-law degree distribution exponent), $m = [2, 4, 6]$ (half of average degree), $T = [0, 0.3, 0.6, 0.9]$ (temperature, inversely related to the clustering coefficient), $N = [100, 500, 1000]$ (network size). For each combination of parameters, 100 networks have been generated. For each network the hyperbolic embedding methods have been executed and the performance has been evaluated using the HD-correlation: Pearson correlation between all the pairwise hyperbolic distances of the nodes in the original network and in the inferred network. The value 1 indicates a perfect correlation and -1 the worst correlation. The plots report for each parameter combination the mean HD-correlation and standard error over the 100 iterations. The following methods are compared: one MCA variant (RA2-MCA1-RAA), the two main coalescent embedding techniques [2] (RA2-ncISO-EA and RA2-ncMCE-EA) and HyperMap-CN [8].

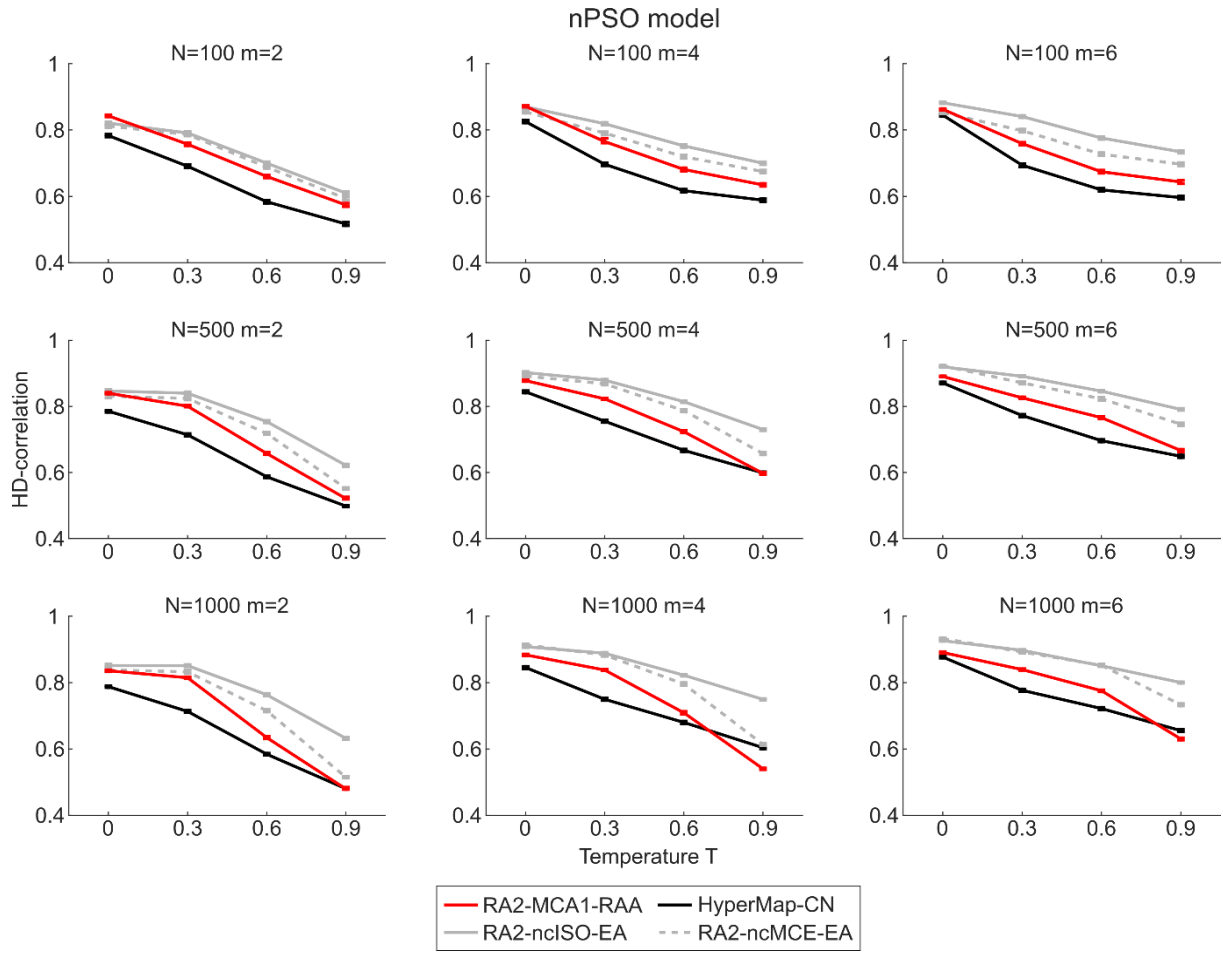

**Suppl. Figure 2. HD-correlation on nPSO synthetic networks.**
Synthetic networks have been generated using the nPSO model with parameters $\gamma = 2.5$ (power-law degree distribution exponent), $m = [2, 4, 6]$ (half of average degree), $T = [0, 0.3, 0.6, 0.9]$ (temperature, inversely related to the clustering coefficient), $N = [100, 500, 1000]$ (network size) and 8 communities. For each combination of parameters, 100 networks have been generated. For each network the hyperbolic embedding methods have been executed and the performance has been evaluated using the HD-correlation: Pearson correlation between all the pairwise hyperbolic distances of the nodes in the original network and in the inferred network. The value 1 indicates a perfect correlation and -1 the worst correlation. The plots report for each parameter combination the mean HD-correlation and standard error over the 100 iterations. The following methods are compared: one MCA variant (RA2-MCA1-RAA), the two main coalescent embedding techniques [2] (RA2-ncISO-EA and RA2-ncMCE-EA) and HyperMap-CN [8].

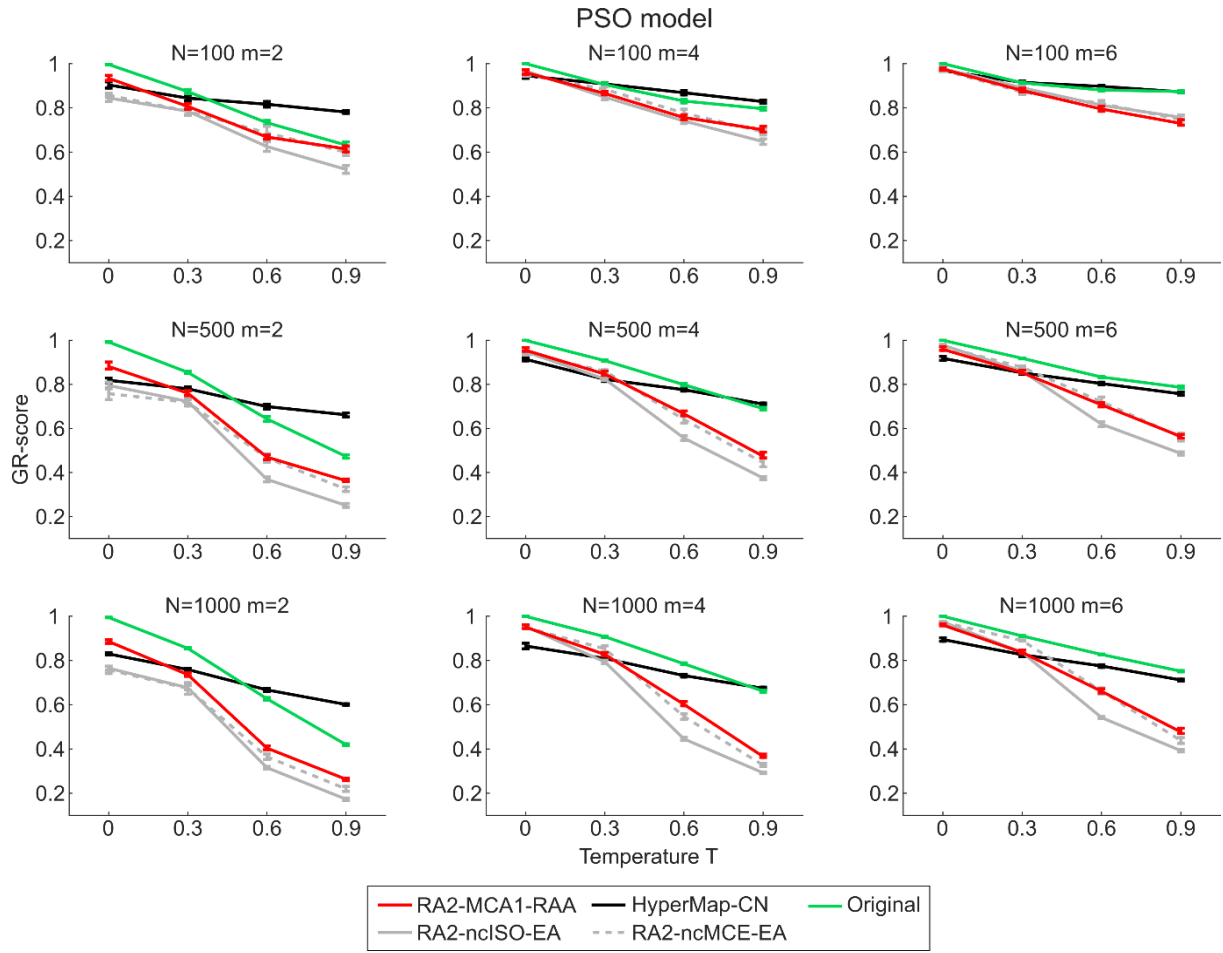

**Suppl. Figure 3. GR-score on PSO synthetic networks.**
Synthetic networks have been generated using the PSO model with parameters $\gamma = 2.5$ (power-law degree distribution exponent), $m = [2, 4, 6]$ (half of average degree), $T = [0, 0.3, 0.6, 0.9]$ (temperature, inversely related to the clustering coefficient), $N = [100, 500, 1000]$ (network size). For each combination of parameters, 10 networks have been generated. For each network the hyperbolic embedding methods have been executed and the performance has been evaluated using the GR-score (details in the Suppl. Methods). The value 1 indicates a perfect greedy routing and 0 represents a completely unsuccessful routing. The plots report for each parameter combination the mean GR-score and standard error over the 10 iterations. The following methods are compared: one MCA variant (RA2-MCA1-RAA), the two main coalescent embedding techniques [2] (RA2-ncISO-EA and RA2-ncMCE-EA) and HyperMap-CN [8]. The GR-score evaluated on the original coordinates of the network is also reported.

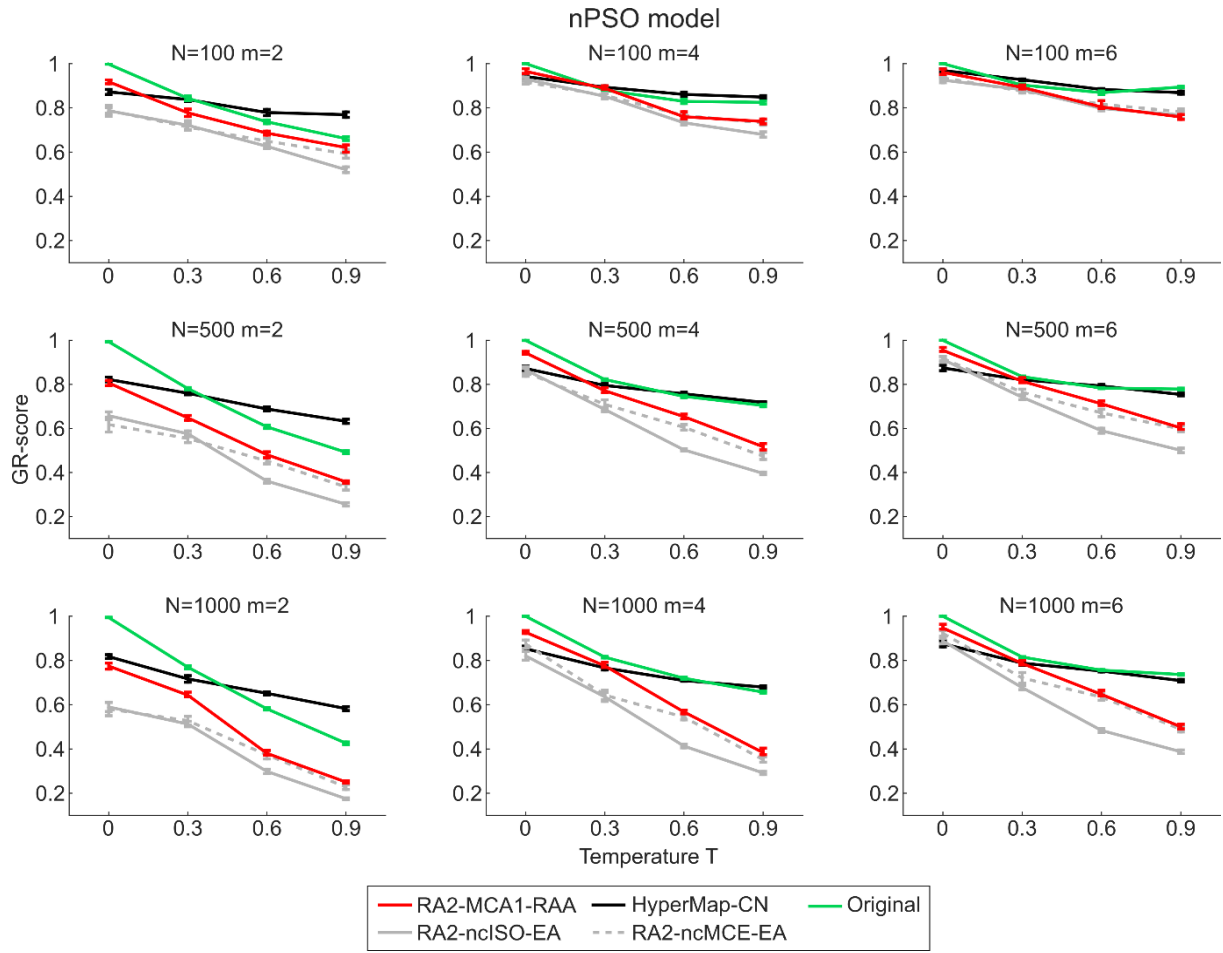

**Suppl. Figure 4. GR-score on nPSO synthetic networks.**
Synthetic networks have been generated using the nPSO model with parameters $\gamma = 2.5$ (power-law degree distribution exponent), $m = [2, 4, 6]$ (half of average degree), $T = [0, 0.3, 0.6, 0.9]$ (temperature, inversely related to the clustering coefficient), $N = [100, 500, 1000]$ (network size) and 8 communities. For each combination of parameters, 10 networks have been generated. For each network the hyperbolic embedding methods have been executed and the performance has been evaluated using the GR-score (details in the Suppl. Methods). The value 1 indicates a perfect greedy routing and 0 represents a completely unsuccessful routing. The plots report for each parameter combination the mean GR-score and standard error over the 10 iterations. The following methods are compared: one MCA variant (RA2-MCA1-RAA), the two main coalescent embedding techniques [2] (RA2-ncISO-EA and RA2-ncMCE-EA) and HyperMap-CN [8]. The GR-score evaluated on the original coordinates of the network is also reported.

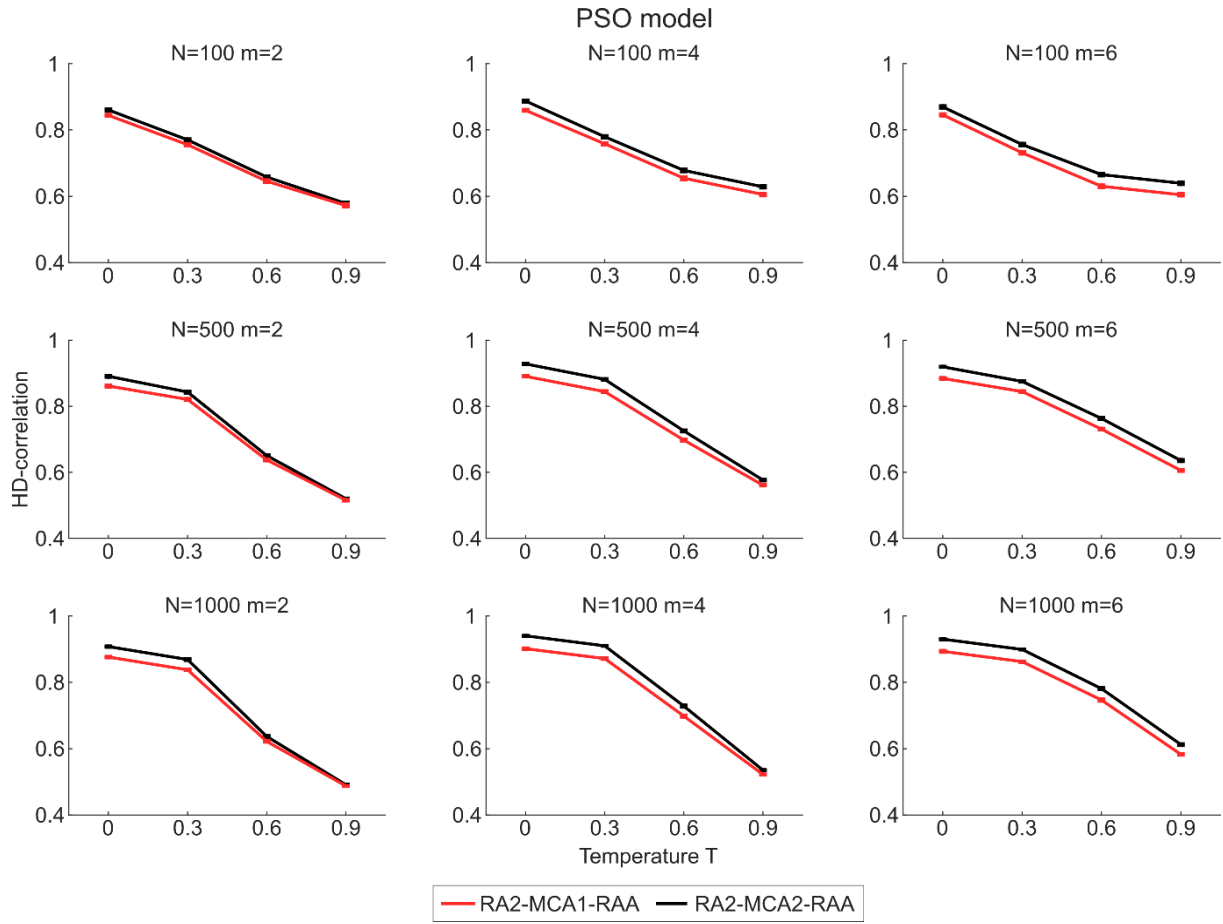

**Suppl. Figure 5. HD-correlation on PSO synthetic networks: MCA1 vs MCA2.**
Synthetic networks have been generated using the PSO model with parameters $\gamma = 2.5$ (power-law degree distribution exponent), $m = [2, 4, 6]$ (half of average degree), $T = [0, 0.3, 0.6, 0.9]$ (temperature, inversely related to the clustering coefficient), $N = [100, 500, 1000]$ (network size). For each combination of parameters, 100 networks have been generated. For each network the hyperbolic embedding methods have been executed and the performance has been evaluated using the HD-correlation: Pearson correlation between all the pairwise hyperbolic distances of the nodes in the original network and in the inferred network. The value 1 indicates a perfect correlation and -1 the worst correlation. The plots report for each parameter combination the mean HD-correlation and standard error over the 100 iterations. The MCA variants RA2-MCA1-RAA and RA2-MCA2-RAA are compared.

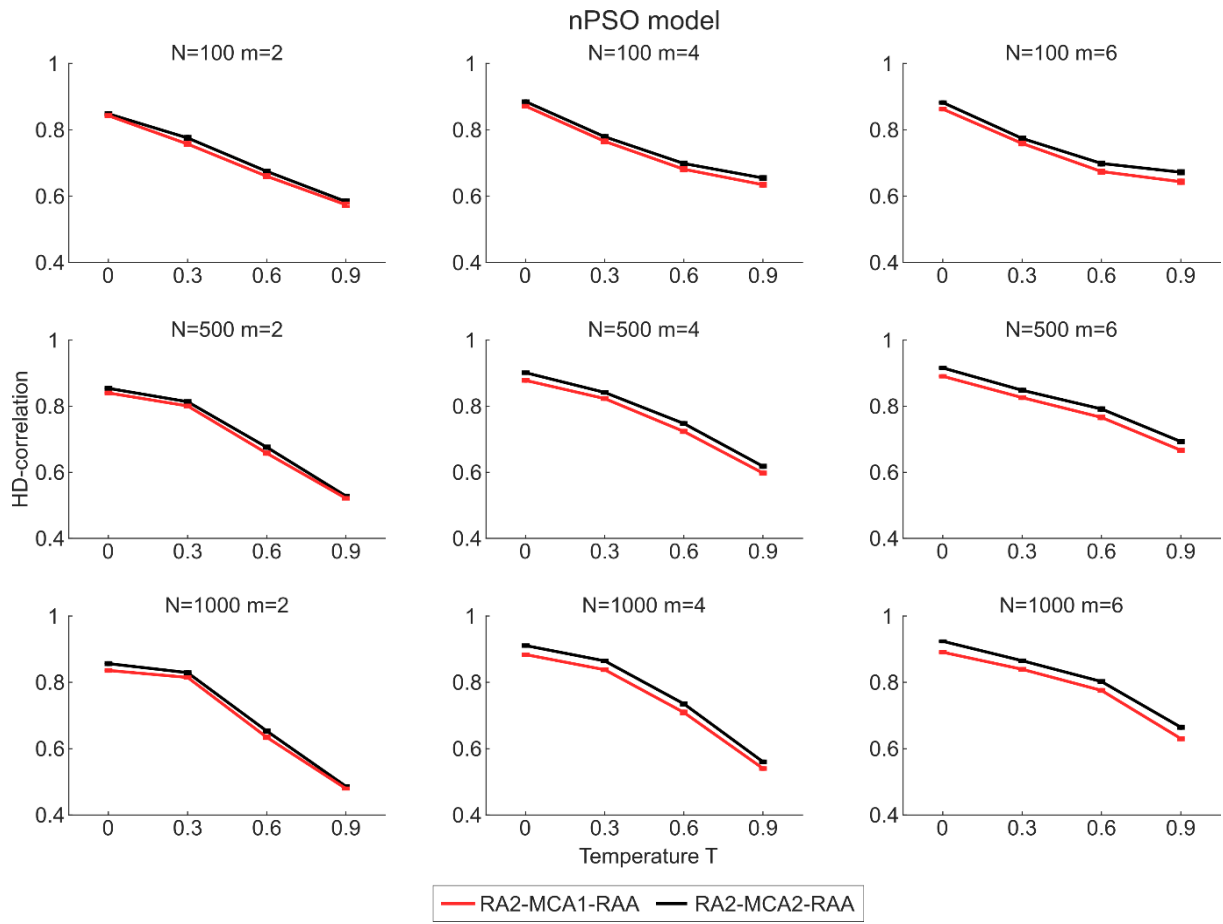

**Suppl. Figure 6. HD-correlation on nPSO synthetic networks: MCA1 vs MCA2.**
Synthetic networks have been generated using the nPSO model with parameters $\gamma = 2.5$ (power-law degree distribution exponent), $m = [2, 4, 6]$ (half of average degree), $T = [0, 0.3, 0.6, 0.9]$ (temperature, inversely related to the clustering coefficient), $N = [100, 500, 1000]$ (network size) and 8 communities. For each combination of parameters, 100 networks have been generated. For each network the hyperbolic embedding methods have been executed and the performance has been evaluated using the HD-correlation: Pearson correlation between all the pairwise hyperbolic distances of the nodes in the original network and in the inferred network. The value 1 indicates a perfect correlation and -1 the worst correlation. The plots report for each parameter combination the mean HD-correlation and standard error over the 100 iterations. The MCA variants RA2-MCA1-RAA and RA2-MCA2-RAA are compared.

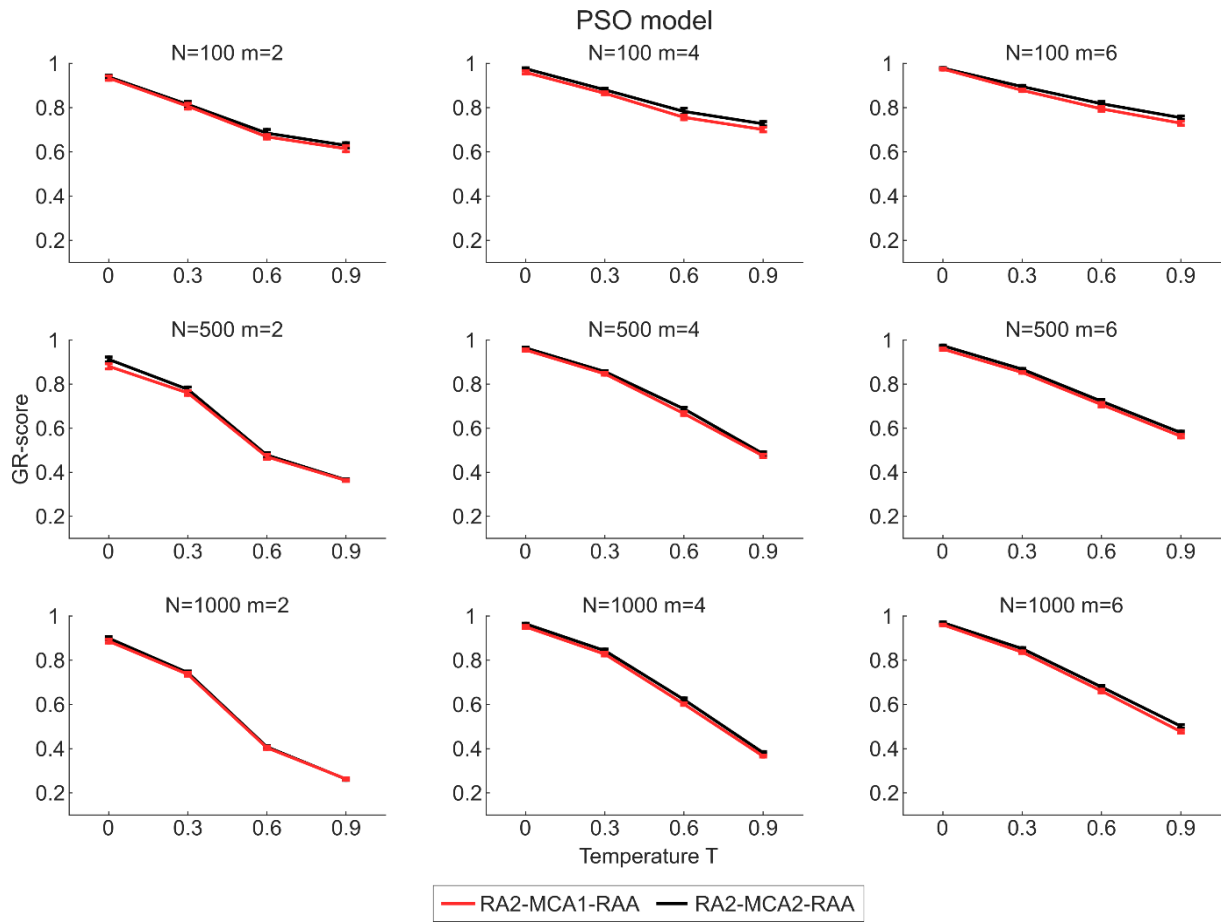

**Suppl. Figure 7. GR-score on PSO synthetic networks: MCA1 vs MCA2.**
Synthetic networks have been generated using the PSO model with parameters $\gamma = 2.5$ (power-law degree distribution exponent), $m = [2, 4, 6]$ (half of average degree), $T = [0, 0.3, 0.6, 0.9]$ (temperature, inversely related to the clustering coefficient), $N = [100, 500, 1000]$ (network size). For each combination of parameters, 10 networks have been generated. For each network the hyperbolic embedding methods have been executed and the performance has been evaluated using the GR-score (details in the Suppl. Methods). The value 1 indicates a perfect greedy routing and 0 represents a completely unsuccessful routing. The plots report for each parameter combination the mean GR-score and standard error over the 10 iterations. The MCA variants RA2-MCA1-RAA and RA2-MCA2-RAA are compared.

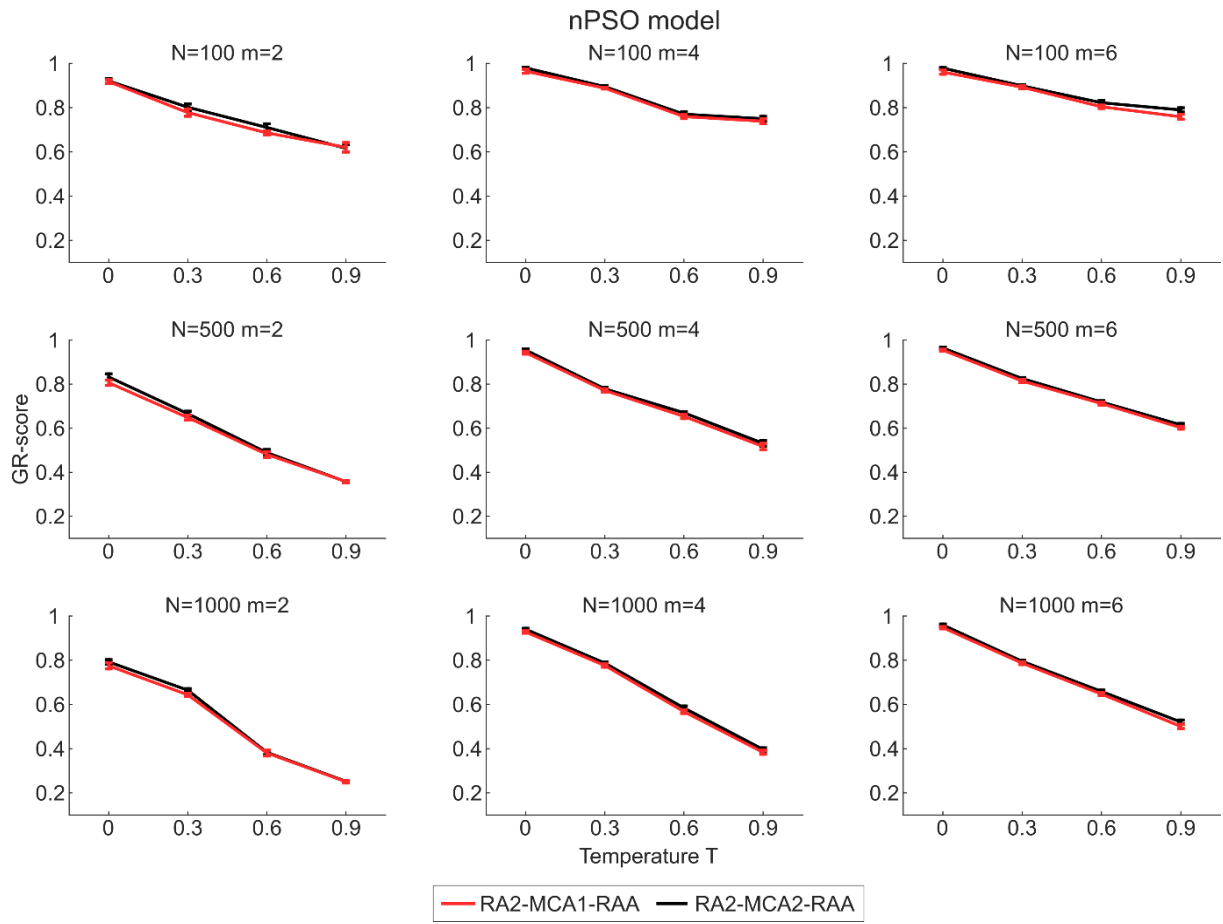

**Suppl. Figure 8. GR-score on nPSO synthetic networks: MCA1 vs MCA2.**
Synthetic networks have been generated using the nPSO model with parameters $\gamma = 2.5$ (power-law degree distribution exponent), $m = [2, 4, 6]$ (half of average degree), $T = [0, 0.3, 0.6, 0.9]$ (temperature, inversely related to the clustering coefficient), $N = [100, 500, 1000]$ (network size) and 8 communities. For each combination of parameters, 10 networks have been generated. For each network the hyperbolic embedding methods have been executed and the performance has been evaluated using the GR-score (details in the Suppl. Methods). The value 1 indicates a perfect greedy routing and 0 represents a completely unsuccessful routing. The plots report for each parameter combination the mean GR-score and standard error over the 10 iterations. The MCA variants RA2-MCA1-RAA and RA2-MCA2-RAA are compared.

|  | RA1 MCA1 RAA HSP | RA1 MCA2 RAA HSP | RA1 MCA2 RAA HD | RA1 MCA1 RAA HD |
| --- | --- | --- | --- | --- |
| Mouse neural | 0.12 | **0.13** | 0.05 | 0.04 |
| Karate | **0.15** | **0.15** | 0.07 | 0.07 |
| Dolphins | **0.15** | 0.14 | 0.10 | 0.10 |
| Macaque neural | **0.60** | 0.58 | 0.36 | 0.35 |
| Polbooks | **0.12** | **0.12** | **0.12** | 0.11 |
| ACM2009 contacts | 0.25 | **0.26** | 0.17 | 0.15 |
| Football | 0.30 | 0.29 | 0.30 | **0.31** |
| Physicians innovation | **0.04** | **0.04** | **0.04** | **0.04** |
| Manufacturing email | **0.41** | **0.41** | 0.20 | 0.16 |
| Littlerock foodweb | 0.26 | **0.28** | 0.07 | 0.08 |
| Jazz | **0.35** | **0.35** | 0.30 | 0.30 |
| Residence hall friends | **0.15** | **0.15** | 0.13 | 0.13 |
| Haggle contacts | **0.56** | 0.54 | 0.44 | 0.44 |
| Worm nervoussys | **0.11** | 0.10 | 0.05 | 0.05 |
| Netsci | 0.21 | 0.19 | 0.30 | **0.31** |
| Infectious contacts | **0.25** | **0.25** | 0.24 | 0.24 |
| Flightmap | 0.54 | **0.56** | 0.50 | 0.46 |
| Email | **0.10** | **0.10** | 0.06 | 0.06 |
| Polblog | **0.15** | **0.15** | 0.12 | 0.12 |
| **mean precision** | **0.25** | **0.25** | 0.19 | 0.19 |
| **mean ranking** | **1.71** | 1.84 | 3.13 | 3.32 |

**Suppl. Table 1. Precision evaluation of link-prediction on small-size real networks: MCA1 vs MCA2.**
For each network 10% of links have been randomly removed (100 realizations), the hyperbolic embedding methods have been executed and the non-observed links in these reduced networks have been ranked either by increasing hyperbolic distance (HD) between the nodes, or by increasing hyperbolic shortest path (HSP), computed as sum of the HD over the shortest path between the nodes [14]. In order to evaluate the performance, the precision is computed as the percentage of removed links among the top-$r$ in the ranking, where $r$ is the total number of links removed. The table reports for each network the mean precision over the 100 iterations, the mean precision over the entire dataset and the mean precision-ranking [34]. For each network the best performance is highlighted in bold. A description of the networks is provided in the Suppl. Datasets section. The MCA variants RA1-MCA1-RAA and RA1-MCA2-RAA are compared.

## Suppl. Algorithm 1. Minimum curvilinear automata (MCA) embedding

INPUT: adjacency matrix $x$
OUTPUT: hyperbolic coordinates $(r, \vartheta)$

1. Pre-weighting using the repulsion-attraction (RA) rule
   a. $x_{ij}^{RA1} = \frac{1+e_i+e_j}{1+CN_{ij}}$
   b. $x_{ij}^{RA2} = \frac{1+e_i+e_j+e_ie_j}{1+CN_{ij}}$
   
   $x_{ij}$ value of $(i,j)$ link in adjacency matrix $x$; $e_i$ external degree of node $i$ (links neither to $CN_{ij}$ nor to $j$); $CN_{ij}$ common neighbours of nodes $i$ and $j$.

2. Prim's algorithm
   I. initialization: sequence $S$ of nodes in the MST (empty); set $V$ of nodes not in the tree (all the nodes).
   II. step $t = 1$, add the highest degree node to $S$; remove it from $V$.
   III. steps $t = 2 \ldots N$, find the edge $e_{ij}$ of minimum weight, where $i \in S$ and $j \in V$; add the node $j$ to $S$; remove it from $V$.
      i. MCA1: $S$ grows in one direction, node $j$ is added always to the same side.
      ii. MCA2: $S$ grows in both directions, node $j$ is added to the side closer to node $i$.

3. Angular coordinates
   a. equidistant adjustment (EA)
   $$\theta_i = \frac{2\pi}{N}(t_i - 1)$$
   $N$ number of nodes in the network; $\theta_i$ angular coordinate of the node $i = 1 \ldots N$; $t_i \in [1, N]$ step in which the node $i$ has been added to the sequence $S$.
   b. repulsion-attraction adjustment (RAA)
   compute the repulsion-attraction between adjacent nodes in $S = s_1, s_2, \ldots s_N$
   $w_1 = RA(s_1, s_2); \ldots ; w_j = RA(s_j, s_{j+1}); \ldots ; w_N = RA(s_N, s_1)$
   normalize the weights so that their sum is $2\pi$
   $$w_j^* = \frac{w_j}{\sum_{k=1}^{N} w_k} * 2\pi$$
   assign the angular coordinates
   $$\theta_i = \sum_{j=1}^{t_i - 1} w_j^*$$

4. Radial coordinates
   Nodes are sorted according to descending degree and the radial coordinate of the $i$-th node in the order is computed according to:
   $$r_i = \frac{2}{\zeta}[\beta \ln i + (1-\beta) \ln N] \quad i = 1, 2, \ldots, N$$
   $\zeta = \sqrt{-K}$, we set $\zeta = 1$; $K$ curvature of the hyperbolic space;
   $\beta = \frac{1}{\gamma - 1}$ popularity fading parameter;
   $\gamma$ exponent of power-law degree distribution, which has been fitted from the network.